\documentstyle[12pt,epsfig]{article}                                
\begin{document}
\titlepage                  
\vspace{0.5in}
\begin{center}
\begin{bf}

 Sensitizing Non-Hermitian Hamiltonian for study of Real Spectra : Controlled 
 broken  PT-symmetry

\vspace{0.1in}

 Biswanath Rath

\vspace{0.1in}
\end{bf}

Department of Physics,North Orissa University , Baripada,Orissa,India

\vspace{0.1in}

E.mail:biswanathrath10@gmail.com

\end{center}
We find that a broken PT-symmetry operator when interacts with suitable
 Hermitian operator ,the new system becomes  completely an un-broken PT-symmetry . Further on
 varying the contribution of Hermiticity, one can delay or control  the broken   PT-symmetry .
 Further beyond the critical contribution , the broken PT-symmetry becomes 
unbroken in nature . Analytical as well as numerical model have been presented.
 In the case of non-analytical model  , we consider the earlier studied
PT-symmetry Hamiltonian : $H=p^{2}+x^{4}+i\lambda x$[C.M.Bender et.al J.Phys A $\bf{34}$
,L.31-36(2001);
 C.R.Handy and X.Q.Wang,J.Phys A $\bf{36}$,11513-11532(2003)]  

\vspace{0.1in}
PACS no- 03.65.Db,11.30.Er

\vspace{0.1in}


\vspace{0.1in}
I.Introduction

\vspace{0.1in}

In classical physics when two systems hardly interact with each other ,which results in conservation of energy[1] . Mathematically if $E_{1,2}$ correspond to 
energy of Hamiltonians $H_{1,2}$ then total energy of the systems is the sum of individual energies i.e
\begin{equation}
E=E_{1}+E_{2}
\end{equation}
However , in quantum mechanics  if two commuting operators $H_{1},H_{2}$ are added  the energy conservation principle is also satisfied[1] i.e 
\begin{equation}
[H_{1},H_{2}]=0
\end{equation}
correspond to 
\begin{equation}
E_{1}+E_{2}=E \rightarrow H=H_{1}+H_{2}
\end{equation}
However if two non-commuting operators 
\begin{equation}
[h_{1},h_{2}]\neq 0
\end{equation}
are added conservation of energy is not preserved i.e
\begin{equation}
\in_{1}+\in_{2}\neq \in \rightarrow h=h_{1}+h_{2}
\end{equation}
This can be verified easily using any two different self-adjoint 
Hamiltonians[2].

\begin{equation}
\Delta \in= \in- \in_{1}-\in_{2}\neq 0
\end{equation}
In above if the two operators are different in nature i.e one is complex and the other is real then the situation becomes different . For example consider
a  complex operator  as 
\begin{equation}
h_{1}=p^{2}-x^{2}+i\sqrt{2}(xp+px)
\end{equation}
having energy eigenvalue [3]
\begin{equation}
\in_{1}=(2n+1)
\end{equation}
and a real self adjoint operator as $h_{2}$ 
\begin{equation}
h_{2}=p^{2}+4x^{2}
\end{equation}
having energy eigenvalue 
\begin{equation}
\in_{2}=2(2n+1)
\end{equation}
Now after mixing ,the new system becomes 
\begin{equation}
h_{1}=2p^{2}+3x^{2}+i\sqrt{2}(xp+px)
\end{equation}
having energy eigenvalue 
\begin{equation}
\in=(2n+1)\sqrt{8}
\end{equation}

Hence the energy conservation principle no longer remains holds good .

\begin{equation}
\Delta \in= \in_{1}-\in_{2} -\in = [3-\sqrt{8}](2n+1)\neq 0
\end{equation}
From the above we conjecture that , non-commuting operators when mix each other ,it is not simple addtion but belongs to quantum interaction in which a new type of operator is generated . In geneal study of this new operator is not an easy . Further if the spectra of generated operator become real ,then it is interesting else reader hardly put interest on new system . In the above model ,the
 complex operator is PT-symmetric in nature . It is to be borne in mind that all PT-invariant operators 
\begin{equation}
[H,PT]=0
\end{equation}
do not yield real eigenvalues . 
Here P stands for the parity operator having the behaviour[4] 
\begin{equation}
P^{-1}xP=-x
\end{equation}
and T stands for the time reversal operator having the behaviour 
\begin{equation}
T^{-1}iT=-i
\end{equation}
As such we can divide PT-symmetry operators mainly in 3 different category as follows .
Real  nature  of eigenvalues[4]
\begin{equation}
H= p^{2} + ix^{3}
\end{equation}
Complex nature of eigenvalues  [4]
\begin{equation}
H= p^{2} + ix
\end{equation}
Mixture of real and complex [5,6] 
\begin{equation}
H= p^{2} + x^{4} + i\lambda x 
\end{equation}
In fact there are many  operators ,which can be formulated  in above three category . Present author feels that operators with  complex eigenvalues are 
studied with less detail compared to operators with completete real spectrum .
 Consideing this , a simple question arises in mind as to converting
 complex to pure real  . In fact it is felt that , if broken PT symmetry 
operator interacts
 with a suitable Hermitian operator , a new unbroken PT-symmetry operator can be
 generated .
There fore  present analysis deals with few examples in analysing unbroken nature of eigenvalues  from Hermitian and bronPT-symmetry operators as follows

 \vspace{0.1in}

II.Matrix model analysis including C-symmetry and P-parity and delaying broken PT-symmetry 

\vspace{0.1in}

Consider a (2x2) PT-symmetry matrix having broken spectra [7,9,10]

\begin{equation}
h^{BPT} =
\left[
\begin{array}{cc}
a+i\sqrt{2}b& b \\
b & a-i\sqrt{2} b \\
\end{array}
\right ]
\end{equation}
The spectra of this model is always complex i.e 
\begin{equation}
\lambda^{BPT}=a\pm ib
\end{equation}

Here we consider a Hermitian operator in the form of (2x2) matrix as 
\begin{equation}
h^{hermitian} =
\left[
\begin{array}{cc}
a&i\sqrt{2}b \\
-i\sqrt{2} b & a \\
\end{array}
\right ]
\end{equation}
having real eigenvalues 
\begin{equation}
\lambda^{hermitian}=a\pm \sqrt{2} b
\end{equation}

Now we add these two operators and produce a new PT-symmetry  operator as[4,7] 
\begin{equation}
H^{UPT} =
\left[
\begin{array}{cc}
2a+i\sqrt{2}b & b(1+i\sqrt{2}) \\
(1-i\sqrt{2})b & 2a-i\sqrt{2}b \\
\end{array}
\right ]
\end{equation}
having real eigenvalues 
\begin{equation}
\lambda^{UPT}=2a\pm  b
\end{equation}
 The corresponding C-symmetry of this new PT-symmetry matrix can be written 
as[7-10] 
\begin{equation}
C =
\left[
\begin{array}{cc}
i\sqrt{2}b & b(1+i\sqrt{2}) \\
(1-i\sqrt{2})b & -i\sqrt{2}b \\
\end{array}
\right ]_{b=1}
\end{equation} 
having eigenvalues 
\begin{equation}
\lambda_{C}=\pm 1
\end{equation}

Similarly P-parity operator is 
\begin{equation}
P =
\left[
\begin{array}{cc}
0 & 1 \\
1 & 0 \\
\end{array}
\right ]
\end{equation} 
having eigenvalues 
\begin{equation}
\lambda_{P}=\pm 1
\end{equation}

Now if one desires to delay the broken symmetry nature of new PT-symmetry ,then then , the  new operator can be written as 
\begin{equation}
H= h^{BPT}+\beta h^{hermitian}
\end{equation}
where $\beta $ is an unknown parameter .Hence the operator is written as 
\begin{equation}
H^{UPT} =
\left[
\begin{array}{cc}
a(1+\beta)+i\sqrt{2}b & b(1+i\sqrt{2}\beta) \\
(1-i\sqrt{2}\beta)b & a(1+\beta)-i\sqrt{2}b \\
\end{array}
\right ]
\end{equation}
having  eigenvalues 
\begin{equation}
\lambda^{hermitian}=a(1+\beta)\pm b \sqrt{2\beta^{2}-1} 
\end{equation}
Hence broken nature of eigevalues will be delayed as long as $\beta \ll \beta_{c}=\frac{1}{\sqrt{2}}$.
.
From the above model , we see that broken-PT-symmetry operator can be converted to un-broken PT-symmetry operator ,when $\beta \gg \beta_{c}$

\vspace{0.1in}

III.Exactly solvable model 

\vspace{0.1in}

It is well known that the operator [4]
\begin{equation}
h^{BPT}=p^{2}+ i \lambda x 
\end{equation}
does not possess any real eigenvalues , eventhough it is PT-symmetry in nature .
On the other hand eigenvalues  of harmonic oscillator 
\begin{equation}
h^{hermitian}=p^{2} + x^{2}
\end{equation}
is purely real[1,2] i.e 
\begin{equation}
E_{n}=2n+1  
\end{equation}

Now we mix these two operator and find a new PT-symmetry operator as 
\begin{equation}
H^{UPT}=2p^{2}+x^{2}+ i\lambda x
\end{equation}
The spectra of this operator is real for any value of $\lambda$ and  is given by[3] 
\begin{equation}
\in_{n}=(2n+1)\sqrt{2}+\frac{\lambda^{2}}{4}
\end{equation}

\vspace{0.1in}

IV:Analytical analysis on numerical model operator 

\vspace{0.1in}

Here we consider the model broken PT-symmetry operator [5,6]
 
\begin{equation}
h^{BPT}=p^{2}+x^{4}+i\lambda x (\lambda \gg 3.169 )
\end{equation}
having broken specta and unbroken spectra provided $\lambda\ll 3.169 $[3,4].
The proposed new PT-symmetry operator can be written as 
\begin{equation}
H= h^{BPT} + \beta h^{hermitian}
\end{equation}
The corresponding Hermitian operator is [2]
\begin{equation}
h^{hermitian}=p^{2}+x^{4}
\end{equation}
Hence the operator H can be written as 
\begin{equation}
H^{UPT}=(1+\beta) p^{2} + (1+\beta)x^{4} + i \lambda x 
\end{equation}
Alternately  
\begin{equation}
H=\frac{H^{UPT}}{(1+\beta)}=p^{2}+x^{4}+i \frac{\lambda}{(1+\beta)} x 
\end{equation}
The above operator will always  yield real eigenvalues for any value of 
$\lambda$ provided $\beta\gg \lambda$ because the ratio $\frac{\lambda}{(1+\beta)}\ll 1$[5,6] . 
There is no restriction on selecting the value of $\lambda $ i.e $\lambda\infty$ .Further no numerical analysis is required as it the said operator has been extensively studied numerically [5,6]. Interested reader can simply use matrix diagonalisation method 
[11] to check the above claim (for one's self satisfaction).Moreover, for a given large value of $\lambda $ if one will choose $\beta $ sufficiently lage , then the eigenvalues of $H^{UPT}$ directly depends on spectraal nature of hermitian operator  $h^{hermitian }$ as  
\begin{equation}
E^{UPT}\sim(1+\beta) E^{hemitian}
\end{equation}
In other words all the eigenvalues of unbroken PT-symmetry operator are real .
Further the above relation implies dominant behaviour of hermiticity over weak
 PT -symmetry contribution ,whose contribution  can be controlled via the parameter .

\vspace{0.1in}

V: Discussions

\vspace{0.1in}

From the above examples , we notice that a broken PT-symmetry operator can  be 
converted to a new un-broken PT-symmetry whem mix with a suitable Hermitian operator . In other words we also conclude that 
\begin{equation}
E_{BPT}+E_{hermitian}\neq E_{UPT}
\end{equation}

This implies that when a complex system mix with a real system ,the new system energy violates the energy conservation principle as well as the nature of
 eigenvalues . Probably a fundamental question in quantum mchanics becomes
 clear . At present it is beyond my limitations to provide a general proof but
one can easily  conjecture that complete real spectra in PT-symmetry operator 
may not far from reality .
In this context ,we also observe that if one adds simple hermitian operator
 like $x^{2,4,6,...}$ , then in some cases one may get pure real spectra ,
 but in some case one may not get a real spectra . For example consider the complete 
broken spectra  Hamiltonian with $\beta x^{2}$ perturbation as 
\begin{equation}
H=p^{2}+i\lambda x + \beta x^{2}
\end{equation}
The eigenvalue of this operator is 
\begin{equation}
E_{n}=(2n+1)\sqrt{\beta}+ \frac{\lambda^{2}}{4\beta}
\end{equation}
For $\beta\rightarrow 0$ energy eigenvalues reflect divergent nature .
Hence simple selection  of operator $\beta x^{2}$ may not be a good choice.
Secodly operator $x^{2}$ as such does not possess any stable 
eigenvalues to focus on energy conservation or violation principle as discussed above .
Lastly author feels that broken PT-symmetry can be controlled as per requirement
Similarly real energy eigenvalues in combined operator can be realised following the above analysis.


\vspace{0.1in}


\vspace{0.1in}


\begin{thebibliography}{99}
\bibitem{Bransden}B.H.Bransden and C.J.Joachain, Quantum Mechanics ,Pearson,Dorling,Kindersley(India),Pvt.Ltd(2000)..
\bibitem{Rath}B.Rath,Ind.J.Phys,$\bf{ B73}$,641(1999)..
\bibitem{Zhang}H.B.Zhang,G.Y.Jiang and G.C.Wang,J.Math.Phys ,$\bf{56}$,70(2017).
\bibitem{Bender}C.M.Bender and S.Boettecher ,Phys.Rev.Lett.$\bf{80}$,5243(1998).
\bibitem{Bender}C.M.Bender ,M.V.Berry,P.N.Meisinger ,Van M.Savage and M.Simsek 
,J.Phys $\bf{A 34}$,L.31-L-36(2001).
\bibitem{Handy} C.R.Handy and X.Q.Wang,J.Phys A $\bf{36}$,11513-11532(2003)]  
\bibitem{Bende}C.M.Bender ,D.C.Brody and H.F.Jones ,Phys.Rev.Lett $\bf{89(27)}$,270401(2002): Erratum $\bf{92(11)}$,119902(2004).
\bibitem{Bender}C.M.Bender ,M.V.Berry and A.Mandilara,J.Phys $\bf{A35}$,L-467(2002).
\bibitem{Rath}B.Rath,The.African Rev.Phys ,$\bf{14:0019}$,139(2019).
\bibitem{Rath}B.Rath,Eur.J.Math.Comp,$\bf{7(1)}$,1(20020).
\bibitem{Rath}P.Mohapapra,B.Rath and P.Mallick,The African Rev.of Phys,$\bf{14:11}$,87(2019).
\end{thebibliography}
\end{document}